\documentclass[reprint, showpacs,
preprintnumbers,
amsmath,amssymb,
aps, floatfix,
superscriptaddress,
prd, nofootinbib]{revtex4-1}

\usepackage{graphicx}
\usepackage{dcolumn}
\usepackage{bm}
\usepackage{hyperref}
\usepackage[mathlines]{lineno}

\usepackage{xcolor}
\usepackage{import}
\usepackage{lipsum}
\usepackage[separate-uncertainty=true,multi-part-units=single]{siunitx}
\usepackage{xfrac}

\usepackage{verbatim}

\newpage

\begin{document}
\title{First Limits on Light Dark Matter Interactions in a Low Threshold Two Channel Athermal Phonon Detector from the TESSERACT Collaboration}

\author{T.K. Bui} \affiliation{International Center for Quantum-field Measurement Systems for Studies of the Universe and Particles (QUP,WPI), High Energy Accelerator Research Organization (KEK), Oho 1-1, Tsukuba, Ibaraki 305-0801, Japan}
\author{C.L. Chang} \affiliation{Argonne National Laboratory, 9700 S Cass Ave, Lemont, IL 60439, USA} \affiliation{Kavli Institute for Cosmological Physics, The University of Chicago, Chicago, IL 60637} \affiliation{Department of Astronomy and Astrophysics, The University of Chicago, Chicago, IL 60637}
\author{Y.-Y. Chang} \affiliation{University of California Berkeley, Department of Physics, Berkeley, CA 94720, USA}
\author{L. Chaplinsky} \affiliation{University of Massachusetts, Amherst Center for Fundamental Interactions and Department of Physics, Amherst, MA 01003-9337 USA}
\author{C.W. Fink} \affiliation{University of California Berkeley, Department of Physics, Berkeley, CA 94720, USA} \affiliation{Now at Los Alamos National Laboratory, Los Alamos, NM 87545}
\author{M. Garcia-Sciveres} \affiliation{Lawrence Berkeley National Laboratory, 1 Cyclotron Rd., Berkeley, CA 94720, USA} \affiliation{International Center for Quantum-field Measurement Systems for Studies of the Universe and Particles (QUP,WPI), High Energy Accelerator Research Organization (KEK), Oho 1-1, Tsukuba, Ibaraki 305-0801, Japan}
\author{W. Guo} \affiliation{Department of Mechanical Engineering, FAMU-FSU College of Engineering, Florida State University, Tallahassee, FL 32310, USA} \affiliation{National High Magnetic Field Laboratory, Tallahassee, FL 32310, USA}
\author{S.A. Hertel} \affiliation{University of Massachusetts, Amherst Center for Fundamental Interactions and Department of Physics, Amherst, MA 01003-9337 USA}
\author{X. Li} 
\email{xinranli@lbl.gov}
\affiliation{Lawrence Berkeley National Laboratory, 1 Cyclotron Rd., Berkeley, CA 94720, USA}
\author{J. Lin} \affiliation{University of California Berkeley, Department of Physics, Berkeley, CA 94720, USA} \affiliation{Lawrence Berkeley National Laboratory, 1 Cyclotron Rd., Berkeley, CA 94720, USA}
\author{M. Lisovenko} \affiliation{Argonne National Laboratory, 9700 S Cass Ave, Lemont, IL 60439, USA}
\author{R. Mahapatra} \affiliation{Texas A\&M University, Department of Physics and Astronomy, College Station, TX 77843-4242, USA}
\author{W. Matava} \affiliation{University of California Berkeley, Department of Physics, Berkeley, CA 94720, USA}
\author{D.N. McKinsey} \affiliation{University of California Berkeley, Department of Physics, Berkeley, CA 94720, USA} \affiliation{Lawrence Berkeley National Laboratory, 1 Cyclotron Rd., Berkeley, CA 94720, USA}
\author{V. Novati} \affiliation{University of Grenoble Alpes, CNRS, Grenoble INP*, LPSC-IN2P3, 38000 Grenoble, France}
\author{P.K. Patel} \affiliation{University of Massachusetts, Amherst Center for Fundamental Interactions and Department of Physics, Amherst, MA 01003-9337 USA}
\author{B. Penning} \affiliation{University of Zurich, Department of Physics, 8057 Zurich, Switzerland}
\author{H.D. Pinckney} \affiliation{University of Massachusetts, Amherst Center for Fundamental Interactions and Department of Physics, Amherst, MA 01003-9337 USA}
\author{M. Platt} \affiliation{Texas A\&M University, Department of Physics and Astronomy, College Station, TX 77843-4242, USA}
\author{M. Pyle} \affiliation{University of California Berkeley, Department of Physics, Berkeley, CA 94720, USA}
\author{Y. Qi} \affiliation{Department of Mechanical Engineering, FAMU-FSU College of Engineering, Florida State University, Tallahassee, FL 32310, USA} \affiliation{National High Magnetic Field Laboratory, Tallahassee, FL 32310, USA}
\author{M. Reed} \affiliation{University of California Berkeley, Department of Physics, Berkeley, CA 94720, USA}
\author{G.R.C Rischbieter} \affiliation{University of Michigan, Randall Laboratory of Physics, Ann Arbor, MI 48109-1040, USA}
\author{R.K. Romani}  \email{rkromani@berkeley.edu} \affiliation{University of California Berkeley, Department of Physics, Berkeley, CA 94720, USA}
\author{B. Sadoulet}\affiliation{University of California Berkeley, Department of Physics, Berkeley, CA 94720, USA}
\author{B. Serfass} \affiliation{University of California Berkeley, Department of Physics, Berkeley, CA 94720, USA}
\author{P. Sorensen} \affiliation{Lawrence Berkeley National Laboratory, 1 Cyclotron Rd., Berkeley, CA 94720, USA}
\author{B. Suerfu} \affiliation{International Center for Quantum-field Measurement Systems for Studies of the Universe and Particles (QUP,WPI), High Energy Accelerator Research Organization (KEK), Oho 1-1, Tsukuba, Ibaraki 305-0801, Japan}
\author{A. Suzuki} \affiliation{Lawrence Berkeley National Laboratory, 1 Cyclotron Rd., Berkeley, CA 94720, USA}
\author{V. Velan} \email{vvelan@lbl.gov} \affiliation{Lawrence Berkeley National Laboratory, 1 Cyclotron Rd., Berkeley, CA 94720, USA}
\author{G. Wang} \affiliation{Argonne National Laboratory, 9700 S Cass Ave, Lemont, IL 60439, USA}
\author{Y. Wang} \affiliation{University of California Berkeley, Department of Physics, Berkeley, CA 94720, USA}
\author{S.L. Watkins} \affiliation{University of California Berkeley, Department of Physics, Berkeley, CA 94720, USA}
\author{M.R. Williams} \email{michaelwilliams@lbl.gov} \affiliation{Lawrence Berkeley National Laboratory, 1 Cyclotron Rd., Berkeley, CA 94720, USA}
\author{J.K. Wuko} \affiliation{University of Massachusetts, Amherst Center for Fundamental Interactions and Department of Physics, Amherst, MA 01003-9337 USA}
\collaboration{TESSERACT Collaboration}
\author{T. Aramaki} \affiliation{Department of Physics, Northeastern University, 360 Huntington Avenue, Boston, MA 02115, USA}
\author{P. Cushman} \affiliation{School of Physics \& Astronomy, University of Minnesota, Minneapolis, MN 55455, USA}
\author{N.N. Gite} \affiliation{University of California Berkeley, Department of Physics, Berkeley, CA 94720, USA}
\author{A. Gupta} \affiliation{University of California Berkeley, Department of Physics, Berkeley, CA 94720, USA}
\author{M.E. Huber} \affiliation{Department of Physics, University of Colorado Denver, Denver, CO 80217, USA} \affiliation{Department of Electrical Engineering, University of Colorado Denver, Denver, CO 80217, USA}
\author{N.A. Kurinsky} \affiliation{SLAC National Accelerator Laboratory/Kavli Institute for Particle Astrophysics and Cosmology, Menlo Park, CA 94025, USA}
\author{B. von Krosigk} \affiliation{Kirchhoff-Institute for Physics, Heidelberg University, 69120 Heidelberg, Germany} \affiliation{Institute for Astroparticle Physics, Karlsruhe Institute of Technology, 76131 Karlsruhe, Germany}
\author{J.S. Mammo} \affiliation{Department of Physics, University of South Dakota, Vermillion, SD 57069}
\author{A.J. Mayer}\affiliation{TRIUMF, Vancouver, BC V6T 2A3, Canada}
\author{J. Nelson} \affiliation{School of Physics \& Astronomy, University of Minnesota, Minneapolis, MN 55455, USA}
\author{S.M. Oser}\affiliation{TRIUMF, Vancouver, BC V6T 2A3, Canada} \affiliation{Department of Physics \& Astronomy, University of British Columbia, Vancouver, BC V6T 1Z1, Canada}
\author{L. Pandey}\affiliation{Department of Physics, University of South Dakota, Vermillion, SD 57069}
\author{A. Pradeep}\affiliation{SLAC National Accelerator Laboratory/Kavli Institute for Particle Astrophysics and Cosmology, Menlo Park, CA 94025, USA}
\author{W. Rau}\affiliation{TRIUMF, Vancouver, BC V6T 2A3, Canada}\affiliation{Department of Physics, Queen’s University, Kingston, Ontario K7L 3N6, Canada}
\author{T. Saab} \affiliation{Department of Physics, University of Florida, Gainesville, FL 32611, USA}
\date{\today}

\begin{abstract}
We present results of a search for spin-independent dark matter-nucleon interactions in a 1 cm$^2$ by 1 mm thick (0.233 gram) high-resolution silicon athermal phonon detector operated above ground. For interactions in the substrate, this detector achieves a  r.m.s.  baseline energy resolution of  $\SI{361.5(4)}{\milli\electronvolt}$, the best for any athermal phonon detector to date. With an exposure of \SI{0.233}{\gram} $\times$ 12 hours, we place the most stringent constraints on dark matter masses between 44 and \SI{87}{\mega\electronvolt\per\square c}, with the lowest unexplored cross section of $\SI{4e-32}{\square\centi\meter}$ at \SI{87}{\mega\electronvolt\per\square c}. We employ a conservative salting technique to reach the lowest dark matter mass ever probed via direct detection experiment. This constraint is enabled by two-channel rejection of low-energy backgrounds that are coupled to individual sensors.
\end{abstract}

\maketitle


\section{\label{sec:introduction}Introduction}

Current and previous generation dark matter (DM) direct detection experiments such as LZ~\cite{LZ:Experiment_2020}, XENONnT~\cite{XenonNT:WS_2023}, DarkSide~\cite{DarkSide-50:2022qzh}, and PandaX~\cite{PandaX4T:SI2023} have largely focused on the detection of weakly interacting massive particles (WIMPs), placing continually tightening constraints on DM models with masses above \SI{1}{\giga\electronvolt\per\square c}. Increasing interest \cite{hochbergMechanismThermalRelic2014, hochbergModelThermalRelic2015, kuflikPhenomenologyELDERDark2017a} in searching for DM below a \si{\giga\electronvolt\per\square c} has motivated some collaborations, including CRESST~\cite{CRESST2019:abdelhameed2019first, CRESST2023}, EDELWEISS~\cite{EDELWEISSLimits}, SuperCDMS~\cite{CPDV1:alkhatib2021light}, DAMIC-M ~\cite{DAMIC-M:arnquist2023first} and SENSEI~\cite{SENSEI2019}, to develop experiments sensitive to sub-GeV DM masses. This mass regime contains models of thermal relic DM (e.g. ELDERs~\cite{Kuflik:2015isi} and SIMPs~\cite{PhysRevLett.113.171301}), freeze-in DM (e.g.  FIMPs~\cite{Hall:2009bx}), secluded DM~\cite{Boehm:2003hm,POSPELOV200853}, Hidden Valleys~\cite{STRASSLER2007374}, asymmetric DM~\cite{PhysRevD.79.115016}, and super-symmetric hidden sectors~\cite{PhysRevLett.101.231301,Hooper_2008} that can either fully or partially produce the correct DM relic abundance. Directly detecting interactions from such DM candidates is challenging, given the eV scale nuclear recoils that they would create in a detector.

Previous efforts to search for nuclear recoils from sub-GeV DM interactions were impeded by an unknown low energy background or backgrounds broadly termed the ``Low Energy Excess'' (LEE)~\cite{EXCESSWorkshop}. These backgrounds are hypothesized to be associated with material effects in the detector environment, such as stress relaxation in the detector holding structure~\cite{AnthonyPetersen2024} or aluminum sensor films~\cite{AlRelaxation}, annihilation of radiation-induced defects in the detector crystal~\cite{DefectLEE}, or scintillation of materials around the detector~\cite{EssigBackgrounds}.

Recently, a subclass of LEE backgrounds has been observed to strongly couple to sensor films from which the detector is constructed~\cite{TwoChannelPaper, CRESSTTwoChannel, TESVeto}, allowing possible discrimination of this background from DM interactions via a coincidence requirement. Multiple athermal phonon readout channels, coupled to the same detector target, would respond simultaneously to athermal phonon bursts from possible DM interactions in the target, while the response to sensor-coupled backgrounds would primarily be confined to a single channel.

In this Letter, we employ this two-channel technique to obtain novel constraints on low-mass DM interactions. This work is part of the TESSERACT (Transition Edge Sensors with Sub-eV Resolution And Cryogenic Targets) initiative, which aims to search for models of sub-GeV DM using a suite of cryogenic detector materials and technologies optimized for low detector thresholds.

\section{\label{sec:Detector}Detector}

\begin{figure}[]
\includegraphics[width=0.9\columnwidth]{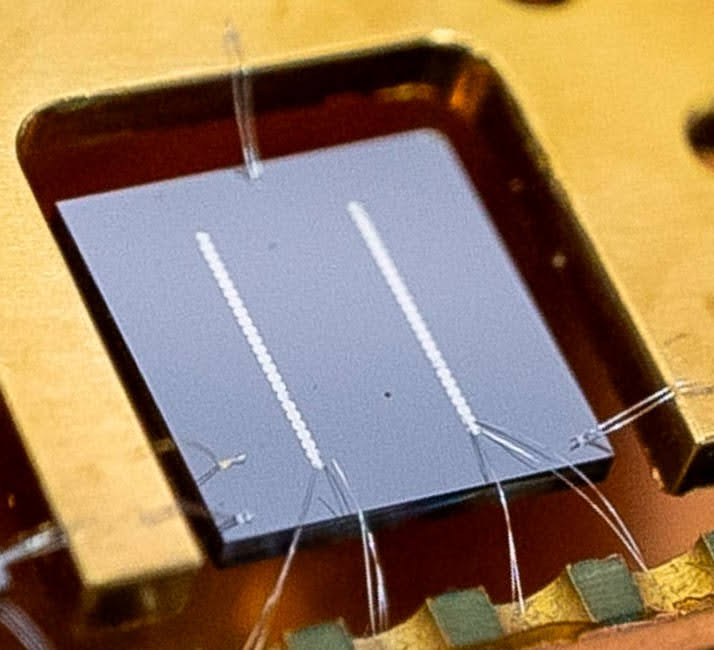}
\caption{The TESSERACT \SI{1}{\square\centi\meter} detector used in this analysis~\cite{photosource}. The detector is supported by wirebonds attached at the top center and bottom corners of the detector. A gold wirebond (left side) is used to cool the detector. The two sensor channels (``left'' and ``right'') can be seen as the parallel lines, each biased and read out separately (see electrical wire bonds to readout PCB). }
\label{fig:device}
\end{figure}

We search for low energy DM interactions in a \SI{1}{\square\centi\meter} by \SI{1}{\milli\meter} thick silicon athermal phonon detector, shown in Fig. \ref{fig:device}, which uses voltage-biased tungsten Transition Edge Sensors (TESs~\cite{IrwinTES1995}, $T_c \approx$ \SI{48}{\milli\kelvin}) coupled to aluminum athermal phonon collection fins in the common Quasiparticle‐trap‐assisted Electrothermal-feedback Transition edge sensor (``QET'') architecture~\cite{irwinQET} to sense phonon bursts in the silicon substrate. We wire 50 QETs into two channels of 25 QETs and read out each channel separately with DC SQUID array amplifiers. This readout scheme was designed to discriminate backgrounds which couple primarily to QET metal films from possible DM interactions. These backgrounds were expected to deposit more energy in one readout channel than in the other, whereas DM interactions with the substrate would produce roughly equal responses in both channels.
Backgrounds associated with stress relaxation in the detector mount were suppressed by suspending the detector by wire bonds as in Ref.~\cite{AnthonyPetersen2024}. We describe low energy background and noise observations in this detector in Ref.~\cite{TwoChannelPaper}.

As we are primarily interested in searching for low energy DM interactions in the detector, we undertook no special precautions to isolate our detector from high-energy backgrounds (e.g. background radioactivity, cosmic rays). We operated our detector in a dilution refrigerator two floors below ground level at the University of California, Berkeley.

To calibrate our detector's response, we injected short ($\sim$\si{\micro\second}) pulses of small numbers of \SI{3.06}{\electronvolt} photons onto the detector, creating athermal pulses of quantized energy~\cite{TwoChannelPaper}. By observing the responses in the two channels, we measure the expected pulse shape for phonon DM-like events, and by combining the responses in the two channels measure a world-leading baseline phonon energy resolution of $\sigma_P = \SI{361.5\pm0.4}{\milli\electronvolt}$ (stat.). 

Current signals from the two detector channels were continuously digitized at \SI{1.25}{\mega\hertz}. Twenty-four hours of data were collected, interleaved with periods where the detector state was monitored to ensure stable operation. To mitigate bias in the analysis, the dataset was divided into 96 15-minute periods. Odd-numbered periods were unblinded and used to develop the analysis pipeline. Even-numbered periods were kept blinded and only opened after the analysis was finalized. Results from the blinded dataset are presented here.

\section{\label{sec:OF} Offline Trigger and Multi-Channel Optimal Filtering}

Offline analysis is performed in two stages: triggering, in which event times are identified, and feature extraction, in which event type and energy are estimated. Both stages use Optimal (Matched) Filtering~\cite{Golwala_Thesis_2000, Watkins_Thesis_2022}, a class of algorithms designed to identify small signal pulses in a noisy data stream.

The main analysis challenge is that the QET response depends on where the energy deposition occurs:
\begin{itemize}
    \item For interactions in the substrate, where we expect $>99$\% of DM recoils to occur, the energy is first carried by athermal phonons. These are absorbed in the Al fins creating quasi-particles, which eventually trap on the W TES sensor, heating it up. This 2-step process leads to a relatively slow rise time. Simultaneously, the $\sim$15\% per-channel phonon collection efficiency suppresses the pulse height. For this class of phonon mediated events, we expect coincident pulses in the two channels with roughly the same amplitude. These events are ``shared'' between both channels. 

    \item In contrast, energy released in the fins of one channel is not coincident with any significant energy in the other channel. These ``singles'' pulses have much sharper rise times and do not suffer from phonon collection inefficiency. The combination of these two effects leads to much higher sensitivity. Our modeling indicates that the left and right channels are more sensitive by a factor $\sim$15 and $\sim$28, respectively.
\end{itemize}
To deal with this challenge,we have extended the classical Optimal Filter method to $N$ simultaneous readout channels and $M$ independent amplitudes.

\subsection{\label{sec:OF_Theory} Building an Optimal Filter}
An Optimal Filter (OF) requires two elements: noise and signal models. The noise is assumed to be stationary (i.e., statistical properties such as moments are independent of time) and Gaussian (Gaussian distribution of Fourier amplitudes at each frequency and random phases, leading to no correlations between the Fourier components). The noise is therefore uniquely characterized by a covariance matrix between all channels at discrete Fourier frequencies. For the signal, at energies well below the saturation of our sensors, the normalized pulse shape (template) is independent of energy~\cite{IrwinTES1995, irwinTransitionEdgeSensors2005, Watkins_Thesis_2022}.

Given the independence of noise at different frequencies, we can construct a $\chi^2$ estimator of signal amplitudes and pulse start times. Minimizing $\chi^2$ is equivalent to maximizing likelihood, so this procedure is optimal for the identification and estimation of energy depositions down to the lowest possible threshold.

In order to simultaneously fit several sensors using a variety of signal shapes, we have designed an ``\mbox{$N \times M$}'' Optimal Filter that simultaneously fits $N$ readout channels with channel-specific pulse shapes scaled by $M$ independent amplitudes (see Appendix~\ref{sec:OF_appendix}). A straightforward example is when $N = M$; each energy deposition has a unique start time, but its amplitudes in the $N$ channels vary independently.

\subsection{\label{sec:OF_Theory} Implementing NxM Optimal Filters}

For substrate events (``shared'' or ``phonon-like''), the best signal-to-noise ratio is obtained by imposing equal pulse amplitudes in both channels, with proper energy normalization of the pulse shapes (templates). This requirement is implemented in a $2 \times 1$ OF. Photon calibration events are used to measure normalized pulse shapes, which are cross checked against the shapes of background shared events and detailed sensor modeling \cite{TESVeto, irwinTransitionEdgeSensors2005, Watkins_Thesis_2022}.

By selecting ``singles'' events, we also empirically generate templates for aluminum fin events; sensor modeling is used to normalize pulse energy. We use a $2 \times 1$ OF to fit singles backgrounds, setting the pulse shape for the non-excited channel to a constant zero waveform. The singles OF uses the non-excited channel to estimate and optimally subtract correlated noise from the excited channel. 

Our data acquisition is continuous (no online trigger). We use a $2\times1$ OF with shared pulse templates to identify events in the recorded data. Events are triggered when the OF amplitude exceeds $4 \sigma_{\text{noise}}$, corresponding to \SI{1.45(2)}{\electronvolt}, where $\sigma_{\text{noise}}$ is the baseline sensor noise. Continuous periods of time when the OF amplitude exceeds the trigger threshold are assembled into discrete events, using a maximum window of 2~ms. More details about the trigger can be found in Appendix~\ref{sec:OF_appendix}.

\begin{figure}[]
\includegraphics[width=1\columnwidth]{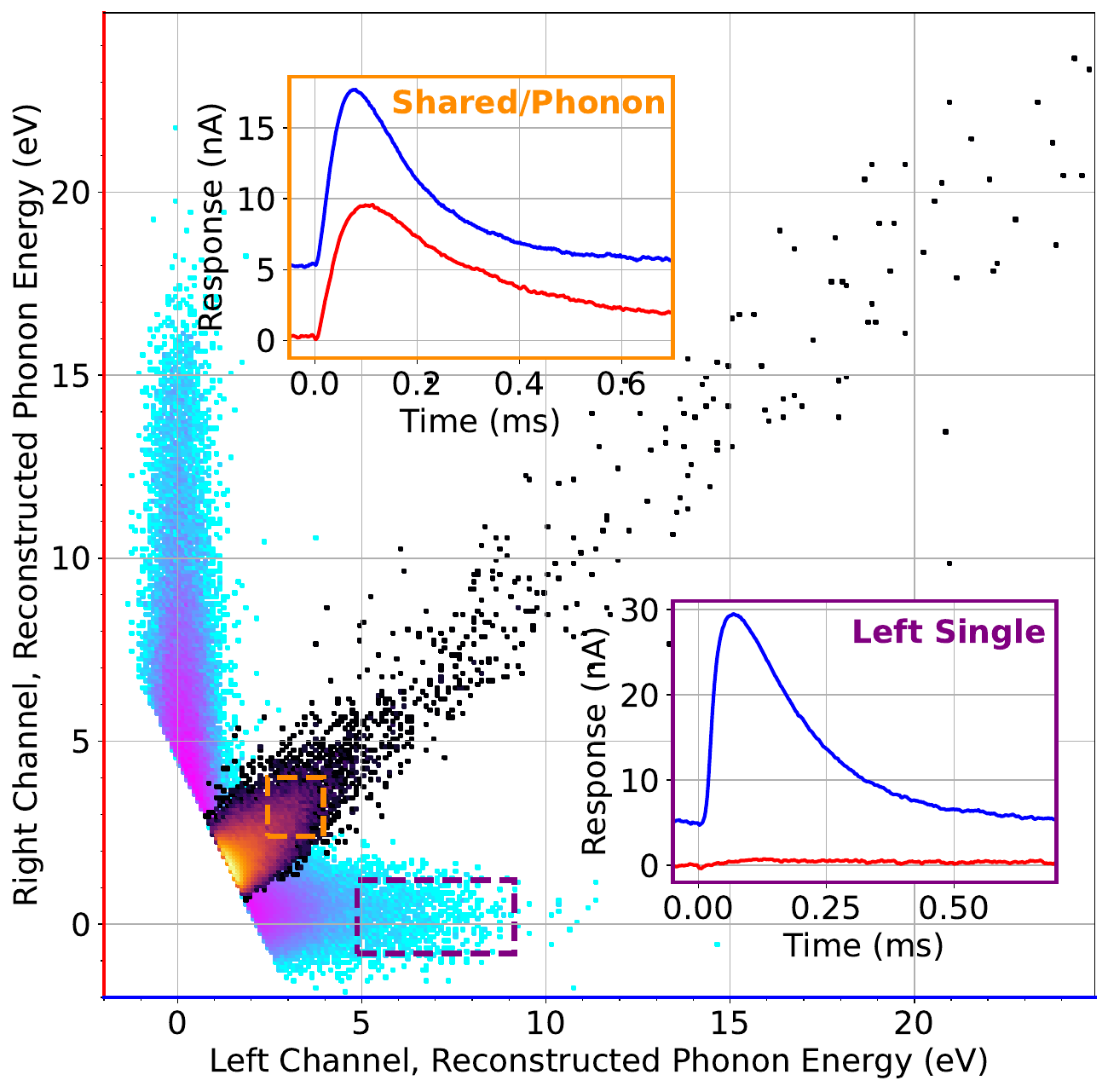}
\caption{(\textbf{Main}) Two dimensional histogram of observed events. Black to orange points on the diagonal show accepted shared/phonon events; blue to purple points show singles events which are inconsistent with a DM response and excluded in the DM search. For energy reconstruction, we use a \mbox{$2 \times 2$} OF and assume a shared pulse shape. As explained in the text this leads an overestimate of the single events, which cluster along the two coordinate axes. 
(\textbf{Top Left Insert}) Averaged shared/phonon event response in the left (blue) and right (red) channels for events in orange dashed box in main figure. (\textbf{Bottom Right Insert}) Averaged left singles event response in the left (blue) and right (red) channels for events in purple dashed box in main figure. In both insert figures, traces are offset and low pass filtered at 50 kHz for clarity.}
\label{fig:2d}
\end{figure}

To extract data features, two types of optimal filters are deployed on each triggered event. First, we inspect energy partitioning between the two readout channels by fitting a \mbox{$2 \times 2$} OF to each triggered event, with shared pulse templates. The best-fit amplitude (i.e.,~energy) in each channel is allowed to vary independently, while maintaining start time coincidence. As shown in Fig.~\ref{fig:2d}, the events divide into three primary populations: shared events consistent with calibrations and DM, on the $x\approx y$ diagonal; left and right channel singles, on the $x$ and $y$ axes. Figure~\ref{fig:2d} also displays the average shared and singles pulse shapes in its inserts. 

To further discriminate potential DM signals (shared events) from background events (singles), we fit three \mbox{$2\times1$} OFs to each event. We assume shared, left channel singles, and right channel singles event topologies, respectively. In this way, each event is associated with three best-fit $\chi^2$ values: $\chi^2_\text{shared}$,  $\chi^2_\text{singleLeft}$, and $\chi^2_\text{singleRight}$. To compare these signal template assumptions, we further define
\begin{eqnarray}
    \delta \chi^2_{SL} = \chi^2_\mathrm{shared} - \chi^2_\mathrm{singleLeft}\\
    \delta \chi^2_{SR} = \chi^2_\mathrm{shared} - \chi^2_\mathrm{singleRight} \\
    \delta \chi^2_{LR} =  \chi^2_\mathrm{singleLeft} - \chi^2_\mathrm{singleRight}.
\end{eqnarray}
\noindent For example,~an event with \mbox{$\delta \chi^2_{SL} > 0$} is more consistent with a singles pulse in the left channel than with a shared pulse, and $\delta \chi^2_{LR} < 0$ is more consistent with a left singles event than a right singles event.

\begin{figure}[]
\includegraphics[width=1\columnwidth]{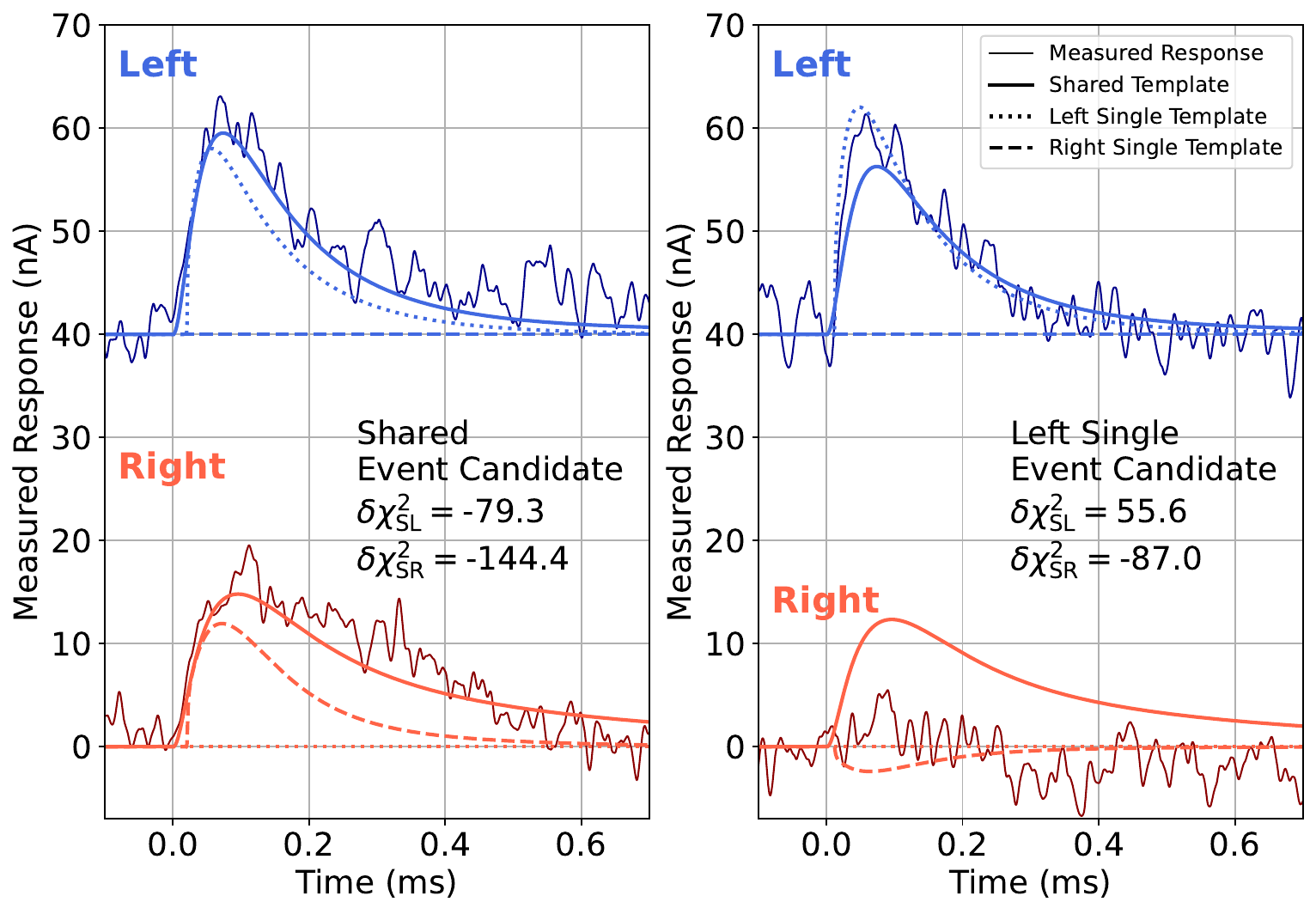}
\caption{A candidate shared event (left panel) and a candidate left singles event (right panel) observed in our detector. We compare the measured response (thin solid line) with three templates: a shared response in both channels (thick solid line), a left singles (dotted), and a right singles (dashed) in the left and right channels, respectively. By using $\delta \chi^2$ statistics, we can determine whether the shape and amplitude in both channels are most consistent with a shared, left singles, or right singles event. Data is filtered with a \SI{50}{\kilo\hertz} low-pass filter and offset for clarity. The left-panel and right-panel events have reconstructed energies of $4.97$ and \SI{3.31}{\electronvolt}, respectively, using a shared template.} 
\label{fig:events}
\end{figure}

The histogram in Fig.~\ref{fig:2d} is colored based on the three \mbox{$2\times1$} OFs---the black-orange heat map represents shared-like events ($\delta \chi^2_{SL} < 0$ and $\delta \chi^2_{SR} < 0$), and the blue-purple heat map represents other events. 
Figure~\ref{fig:events} depicts two example measured events, compared with best-fit shared and singles templates.

\section{\label{sec:Analysis}Event Selection and Efficiencies}

Our trigger finds 434,090 events, each reconstructed with a trace length of 20 ms. To remove events in periods of poor data quality where the detector response will be abnormal (e.g. periods of high electromagnetic interference (EMI) or vibration induced noise, or ``pileup'' events with multiple triggers in the same OF window), we impose data quality cuts. To reject periods of high EMI or noise, as well as elevated detector temperature following e.g. high-energy cosmic ray interactions, we cut events where the pre-pulse quiescent current (baseline) or difference between pre- and post-pulse currents (slope) fell outside a predetermined range from the unblinded dataset. These  cuts remove \SI{4.5}{\percent} of events. We additionally perform a cut based on the $\chi^2$ value extracted from the OF for every given pulse (considering only frequencies below 50 kHz and assuming a phonon template), rejecting events which are inconsistent with the expected pulse shape. This $5.6 \sigma$ low frequency $\chi^2$ cut has the effect of removing pileup, triggers at incorrect times, and events with high noise or abnormal pulse shape. Events passing these cuts in both of our detector channels are preliminarily accepted as DM candidate events.

\begin{figure}[t!]
\includegraphics[width=1\columnwidth]{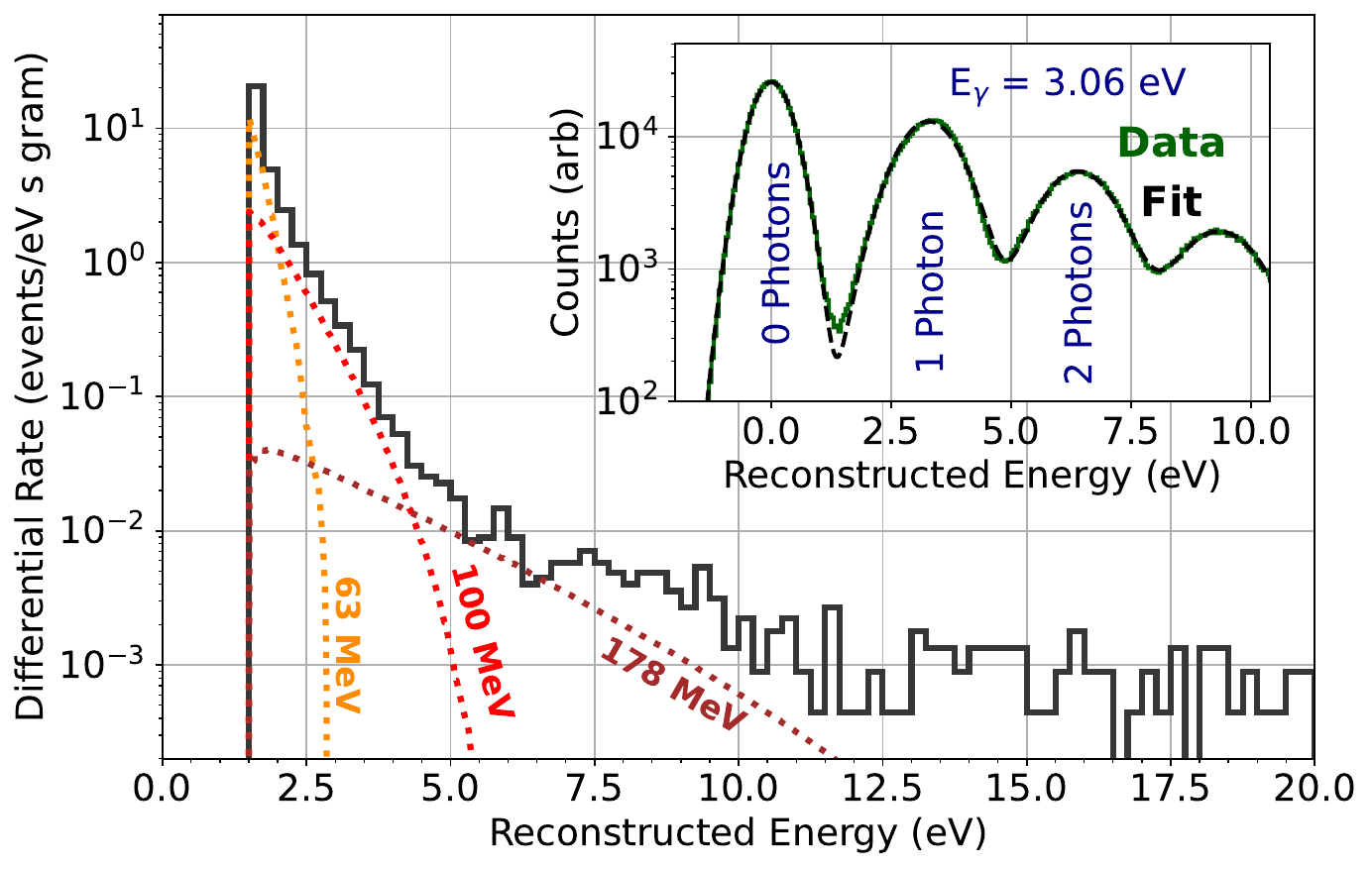}
\caption{(Main) Spectrum of observed events passing quality cuts and shared event cuts (black). Dotted lines show the modeled DM interaction spectra at the limit of exclusion for DM with masses $63$, $100$, and $\SI{178}{\mega\electronvolt/\square c}$. (Insert) Calibration spectrum, showing peaks from absorbing zero, one, two, etc. photons.}
\label{fig:spectrum}
\end{figure}

To reject backgrounds which couple primarily to the phonon sensors, we perform a final selection of events with $\delta \chi^2_{SL} < 0$ and $\delta \chi^2_{SR} < 0$ (i.e. events that are more shared-like than singles-like). Events which pass this additional $\delta \chi^2$ cut are accepted into our final DM analysis. We lastly place an analysis threshold of \SI{1.5}{\electronvolt}, removing any events below this energy. 
The final event spectrum can be seen in Fig.~\ref{fig:spectrum}.

We measure the impact of these cuts on the detector nuclear recoil (NR) efficiency using injected virtual pulses with an ideal shared pulse shape (``salt")~\cite{salting:li2024modelingdifferentialratesignal}. Salt pulses at a range of NR energies, as low as \SI{0.361}{\electronvolt}, are injected into the continuous data stream pre-trigger. For each injected energy, we measure the binned difference between the salted and unsalted spectrum and normalize by the total number of salts. This net differential signal response estimates inefficiencies due to triggering, analysis cuts, and measures energy smearing. The response to sub-threshold recoil energies observed through \textit{noise boosting}~\cite{salting:li2024modelingdifferentialratesignal} is also studied provided that the signal response is linear and has small true signal pileup. See Appendix~\ref{sec:salting_appendix} for details. 
This is convolved with the standard spin-independent NR DM spectrum $dR/dE$~\cite{DM_parameters:LEWIN199687_DM_density} to build the model of measured spectrum $dR/dE'$. The $dR/dE'$ spectra for DM masses of \SI{63}, \SI{100}, and \SI{178} {\mega\electronvolt\per\square c} can be seen overlaid on the measured event spectrum in Fig.~\ref{fig:spectrum}.

\section{\label{sec:results}Results}
The total exposure is $\SI{0.233}{\gram}\times \SI{12}{\hour}$. The interaction rate between DM and silicon nuclei is modeled assuming scattering from DM with a local density of \mbox{0.3~GeV/cm$^3$} and a Maxwell–Boltzmann velocity distribution, using the parametrization of Lewin and Smith~\cite{DM_parameters:LEWIN199687_DM_density}. This model is convolved with the detector net differential sensitivity estimated via salting to get the net differential response for a given DM model~\cite{salting:li2024modelingdifferentialratesignal}.

\begin{figure}[t!]
\includegraphics[width=1\columnwidth]{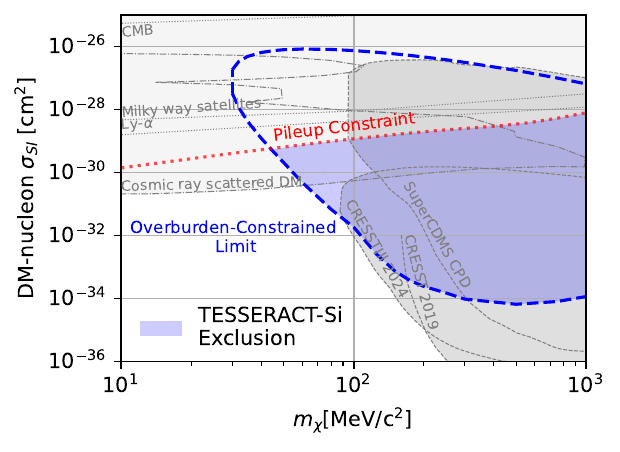}
\caption{The \SI{90}{\percent} C.L. limits on spin-independent DM below \SI{1}{\giga\electronvolt\per\square c}. The blue shaded region shows the exclusion from this work. The blue dashes line represents the exclusion from the Yellin optimum interval test when considering the effect of the overburden. The red dotted line is from the requirement that there be no DM signal pileup. Previously excluded DM phase space from the CRESST~\cite{CRESST:2024cpr,CRESST2019:abdelhameed2019first} and SuperCDMS~\cite{CPDV1:alkhatib2021light} collaborations are shown in gray. Constraints from cosmology and astrophysics are shown in light gray~\cite{cappiello2019strong, bringmann2019novel, DES:2020fxi, PhysRevLett.128.171301,PhysRevLett.121.081301}. }
\label{fig:limit}
\end{figure}

The limit on the DM-nucleon spin-independent scattering cross-section $\sigma_{SI}$ is calculated with the Yellin optimum interval method~\cite{Yellin_Optimum_Interval_2002, Yellin_Optimum_Interval_2007} at the \SI{90}{\percent} confidence level, using reconstructed energy as the only observable. The background is completely un-modeled and is thus considered potential DM signal. As is standard for DM searches at relatively high cross-sections~\cite{CPDV1:alkhatib2021light,CRESST:2024cpr}, we take into consideration the shielding effect from the overburden. This includes the atmosphere and \SI{3}{\meter} concrete floors above the lab, and is calculated using the \texttt{Verne} code base~\cite{Kavanagh_2018, Verne_Software}. The resulting limit given this overburden consideration is seen as the blue dashed line in Fig.~\ref{fig:limit}. 

Some of this nominally excluded parameter space has a DM interaction rate that would lead to significant signal pileup. In this regime, the $dR/dE'$ does not scale linearly with $\sigma_{SI}$, rendering any linear interaction model incorrect~\cite{salting:li2024modelingdifferentialratesignal}. To account for this, we further restrict the exclusion region to include only DM masses and cross-sections that would not produce significant pileup. To estimate the rate at which we begin to see pileup effects, we note that 91~$\mu$s after a phonon pulse is triggered, the OF will relax to 10\% of its peak amplitude. This suggests that a pileup rate of $1/(2\times91~\mu s)$= 5.5~kHz is a reasonable upper exclusion boundary, below which our linear differential rate modeling is valid.  This is seen as the red dotted curve in Fig.~\ref{fig:limit}. Future analyses can be designed specifically to search for DM with high pileup rates. These analyses will likely extend the reaches of this detector and others to substantially lower DM masses and higher interaction cross sections ~\cite{Das:2022srn, NoiseLimit-work-in-progress}.

New bounds on cross sections as low as \SI{4e-32}{\square\centi\meter} below \SI{87}{\mega\electronvolt\per\square c} DM mass are established. We place constraints on DM-nucleon cross sections down to masses of \SI{44}{\mega\electronvolt\per\square c} at \SI{4.67e-30}{\square\centi\meter}, the lowest mass ever probed by a particle-like DM search, as a consequence of our excellent energy resolution. We place our most stringent bounds for DM with a cross-section of \SI{6.56e-35}{\square\centi\meter} at \SI{500}{\mega\electronvolt\per\square c}. The full DM exclusion region from this work can be seen as the blue shaded region in Fig.~\ref{fig:limit}.

\section{\label{sec:conclusion}Conclusion}

\noindent This letter presents the lowest-mass sensitivity of any dark matter nuclear recoil search, as a result of our unparalleled energy resolution of \SI{361.5\pm0.4}{\milli\electronvolt}. By treating our unknown background with the optimum interval, we are able to place world-leading constraints on DM between \SI{44}{\mega\electronvolt\per\square c} and \SI{87}{\mega\electronvolt\per\square c}. These results show the potential of low-threshold superconducting sensors for exploring new DM parameter space. Through both the direct minimization of low energy excess backgrounds and the discrimination of these backgrounds with novel cryogenic targets like gallium arsenide and superfluid He ~\cite{PhysRevD.100.092007,SPICE:2023tru}, TESSERACT aims to substantially improve upon this surface DM search, as well as search for other DM interactions. 

\begin{acknowledgments}

The athermal phonon detector used in this search builds upon the work that has been done by the CDMS and SuperCDMS collaborations \cite{irwinQET,Yen:apl14:QPtransport, Pyle:2012hma}.
This work was supported in part by DOE Grants  DE-SC0022916, DE-SC0019319, DE-SC0025523 and DOE Quantum Information Science Enabled Discovery (QuantISED) for High Energy Physics (KA2401032) and National Science Foundation Grants 2111375 and 1106400.
This material is based upon work supported by the Department of Energy National Nuclear Security Administration through the Nuclear Science and Security Consortium under Award Number(s) DE-NA0003180 and/or DE-NA0000979. Work at Lawrence Berkeley National Laboratory was supported by the U.S. DOE, Office of High Energy Physics, under Contract No. DEAC02-05CH11231. Work at Argonne is supported by the U.S. DOE, Office of High Energy Physics, under Contract No. DE-AC02-06CH11357. Work at Texas A\&M University was supported by the Mitchell Institute and by the U.S. DOE, Office of High Energy Physics, under Contract Nos. DE-SC0018981 and DE-SC0021051. W.G. and Y.Q. acknowledge the support by the National High Magnetic Field Laboratory at Florida State University, which is supported by the National Science Foundation Cooperative Agreement No. DMR-2128556 and the state of Florida.
\end{acknowledgments}

\appendix

\section{\label{sec:OF_appendix}Multi-Channel Optimal Filtering}

Our offline analysis trigger extends the Optimal (Matched) Filter algorithm~\cite{Golwala_Thesis_2000, Watkins_Thesis_2022} to $N$ simultaneous readout channels and $M$ potential signal shapes. It is based on $\chi^2_{\text{s}} (\vec{A}, t_s)$, defined below, which represents the consistency of an observed trace with the hypothesis $s$ under consideration.

\begingroup\makeatletter\def\f@size{9.5}\check@mathfonts
\def\maketag@@@#1{\hbox{\m@th\normalfont#1}}
\begin{equation}
\begin{gathered}
\chi^2_{\text{s}}(\vec{A},t_s) \equiv \int{df \chi^2_{\text{s}}(\vec{A},t_s,f)} \:\: , \\
\text{and } \chi^2_{\text{s}}(\vec{A},t_s,f) \equiv 
\sum_{i,j=1}^N \left(v_i^*(f) - \sum_{\alpha=1}^M A_\alpha e^{2i \pi f t_s} s_{i \alpha}^*(f) \right) \times \\C^{-1}_{ij}(f) \left(v_j(f) - \sum_{\beta=1}^M A_\beta e^{- 2i \pi f t_s} s_{j \beta}(f) \right).\\
\end{gathered}
\label{eq:chi2_full}
\end{equation}\endgroup

\noindent Here, $\vec{A}$ is an $M$-length real vector of signal template amplitudes; $t_s$ is the common starting time of pulses; $v_i(f)$ is the Fourier transform of the observed trace in channel $i$; $s_{i\alpha}(f)$ is the Fourier transform of the signal template for channel $i$ corresponding to amplitude $A_{\alpha}$;  $C(f)$ is the noise covariance matrix (i.e., the noise cross-spectral density between pairs of channels); $C^{-1}_{ij}(f)$ is the $(i,j)$ element of $C(f)^{-1}$; and we integrate over frequencies. 

The Optimal Filter estimates of $\vec{A}$ and $t_s$ are obtained by minimizing $\chi^2_{\text{s}}(\vec{A},t_s)$.
In practice, the time traces are digitized at a sampling frequency $f_{\text{samp}}$ and a discrete Fourier transform is performed over a trace time-length of $T$. The integral then becomes a sum over frequencies between $-f_{\text{samp}}/2$ and $f_{\text{samp}}/2$ in steps of $1/T$. In this study, $f_{\text{samp}}=1.25$~MHz and $T=20$~ms.

\textbf{Construction of an optimal filter trigger.} In order to identify potential events, we let a \mbox{$2 \times 1$} OF with shared templates and a common amplitude run freely along the time sequence of the data stream.   
We then construct $\Delta \chi^2_{s}(t)$ as below, where $\hat{A}$ is the value of $\vec{A}$ that minimizes $\chi^2_{s} (\vec{A}, t)$:
\begin{equation}
\Delta \chi^2_{s}(t) \equiv  \chi^2_{s}(0,t) -  \chi^2_{s}(\hat{A},t)
\label{eq:delta_chi2_definition}
\end{equation}

\noindent This represents, for a given trace, the difference between the minimal $\chi^2$ in Fourier space and the $\chi^2$ that would be obtained by assuming no signal. Our $N \times M$ Optimal Filter trigger runs on $\Delta \chi^2_{s}(t)$, optimally selecting pulses that cross a certain threshold. This allows us to cleanly vary the signal template without changing the trigger parameters. It can be shown that $\Delta \chi^2_{s}(t)$ is quadratically dependent on the best-fit signal amplitude $\hat{A}$. For the simple case of one readout channel and one signal template, this trigger is thus identical to triggering on the Optimal Filter amplitude, as described in \cite{Watkins_Thesis_2022}.

Events are selected when $\Delta \chi^2_\text{shared}$ exceeds 16, or equivalently when the best-fit amplitude $\hat{A}$ exceeds $4 \sigma_{\text{base}}$. Here, $\sigma_{\text{base}}$ is the baseline resolution of $\hat{A}$, i.e.,~the magnitude of fluctuations of $\hat{A}$ due only to noise.

For a given pulse, $\Delta \chi^2_\text{shared}$ will exceed the trigger threshold continuously for some amount of time, so we select triggers by assembling above-threshold regions of time within a 2-ms window into discrete events. This window was selected to capture all unsaturated signals while minimizing deadtime. Within each region, the trigger is placed at the time that maximizes $\Delta \chi^2_\text{shared}$.

 \textbf{Analysis optimal filters.} We then implement OFs on triggered events to extract pulse features. As the template shape varies, the pulse start time is also allowed to vary within the 2~ms time window. For instance, to discriminate potential DM signals (shared events) from background events (singles), we fit three \mbox{$2\times1$} OFs to each event and construct $\delta \chi^2$, as explained in the main text. (Note that  $\delta \chi^2$ is different from $\Delta \chi^2_\text{shared}$.) Cuts based on $\delta \chi^2$ are, in essence, maximum likelihood ratio tests on the data. 

\section{\label{sec:salting_appendix}Noise modeling with the Salting Method}
The upward fluctuation of noise allows additional sensitivity to events with sub-threshold true energies that are boosted above the trigger threshold. This idea has been widely adopted in DM searches \cite{CPDV1:alkhatib2021light, CRESST2019:abdelhameed2019first}. The upward fluctuation is usually limited to $3\sigma$ of the baseline energy resolution to prevent sensitivities to zero energy events. Discussions in \cite{salting:li2024modelingdifferentialratesignal} pointed out that a more rigorous modeling of the noise smearing the true event energy without the arbitrary $3\sigma$ cut-off can be achieved by salting the raw data traces with ideal signal pulses.

The core concept is the net differential response, defined as
\begin{equation}
    \Delta f(E'|E) \equiv f(E'|E) - f(E'|0)
\end{equation}
where $E$ is the true energy and $E'$ is the measured energy. The quantity $f(E'|E)$ represents the probability distribution of $E'$ given that events with energy $E$ are present. The $- f(E'|0)$ term accounts for the reduction of noise-only time periods in the measurement as signals are added. In previous works, this term is ignored because the overlap between $f(E'|E)$ and $f(E'|0)$ above the trigger threshold is negligible. However, it is problematic if signals with $E\sim0$ are considered, resulting in underestimation of DM cross sections at low-masses.

We estimate $\Delta f(E'|E)$ using the salting method. First, $N_s$ pulses (salts) with ideal signal pulse shape and energy $E_s$ are injected into the continuously recorded raw traces of total exposure time $T$ randomly in time. The random injection times are separated by the triggered trace length to prevent pileup of two salt pulses. Second, the salted traces are processed by the offline trigger and filtering algorithm as described in Sec.~\ref{sec:OF}, then selected under the same criteria as in Sec.~\ref{sec:Analysis}. The exact same analysis is done for both salted and unsalted datasets to ensure the accurate measurement of trigger and event selection efficiencies. Then, the measured spectrum $\widehat{\frac{dR}{dE'}}(E'|S+s)$ is compared with the one before salting $\widehat{\frac{dR}{dE'}}(E'|S)$, and normalized by $r_s=N_s/T$ to estimate  $\Delta f(E'|E)$
\begin{equation}
    \widehat{\Delta f}(E'|S+E_s) = \frac{\widehat{\frac{dR}{dE'}}(E'|S+s) - \widehat{\frac{dR}{dE'}}(E'|S)}{r_s}
\end{equation}
where $S$ represent the potential DM signal in the unsalted spectrum, and $s$ represents the salt. Finally, salting is repeated at energies from \SI{0.361}{\electronvolt} to \SI{30}{\electronvolt}. The limit of DM with mass $m_\chi$ is calculated with the Yellin optimum interval method \cite{Yellin_Optimum_Interval_2007}, with 

\begin{equation}
\begin{split}
    &\Delta \frac{dR}{dE'}(E'|S,s(m_\chi,\sigma_0)) \\
    &\equiv
    \int_0^\infty \frac{dR}{dE}(E_s|s(m_\chi,\sigma_0))\widehat{\Delta f}(E'|S+E_s) dE_s
    \end{split}
\end{equation}
scaling with $\sigma_\mathrm{SI}$ but not $\frac{dR}{dE'}$. Here $\sigma_0$ is the reference DM cross section. See \cite{salting:li2024modelingdifferentialratesignal} for full discussion. 

\bibliography{references}

\appendix

\end{document}